\documentclass[aip,jap,reprint]{revtex4-1}

\usepackage{graphicx}
\usepackage{dcolumn}
\usepackage{bm}
\usepackage{verbatim}
\usepackage{color}

\begin{document}


\title{Imaging the formation of a p-n junction in a suspended carbon nanotube with scanning photocurrent microscopy}

\author{Gilles Buchs}
 \email{g.buchs@tudelft.nl}
 \affiliation{Kavli Institute of Nanoscience, TU-Delft, Post Office Box 5046, 2600 GA Delft, The Netherlands}%

\author{Maria Barkelid}
 \affiliation{Kavli Institute of Nanoscience, TU-Delft, Post Office Box 5046, 2600 GA Delft, The Netherlands}%

\author{Salvatore Bagiante}
 \affiliation{Istituto per la Microelettronica e Microsistemi, Consiglio Nazionale
delle Ricerche, Stradale Primosole 50, I-95121 Catania, Italy}

\author{Gary A. Steele}
 \affiliation{Kavli Institute of Nanoscience, TU-Delft, Post Office Box 5046, 2600 GA Delft, The Netherlands}%

\author{Val Zwiller}
 \affiliation{Kavli Institute of Nanoscience, TU-Delft, Post Office Box 5046, 2600 GA Delft, The Netherlands}%


\begin{abstract}
We use scanning photocurrent microscopy (SPCM) to investigate individual suspended semiconducting carbon nanotube devices where the potential profile is engineered by means of local gates. In situ tunable p-n junctions can be generated at any position along the nanotube axis. Combining SPCM with transport measurements allows a detailed microscopic study of the evolution of the band profiles as a function of the gates voltage. Here we study the emergence of a p-n and a n-p junctions out of a n-type transistor channel using two local gates. In both cases the $I-V$ curves recorded for gate configurations corresponding to the formation of the p-n or n-p junction in the SPCM measurements reveal a clear transition from resistive to rectification regimes. The rectification curves can be fitted well to the Shockley diode model with a series resistor and reveal a clear ideal diode behavior.
\end{abstract}

\maketitle

\section{INTRODUCTION}

The unique electronic properties of carbon nanotubes make them ideal systems for future large-scale integrated nanoelectronics circuits \cite{Review_CNT07}. Due to their quasi-one-dimensional geometry, the electronic bands of carbon nanotubes can be engineered by means of electrostatic doping. In this context, p-n junction diodes \cite{Lee_APL_04,Lee_APL_05,Bosnick_APL_06,Gabor_Science_09,Mueller_2010,Michigan_11} as well as tunable double quantum dots working in the single particle regime have been realized in suspended nanotube devices using local gates \cite{Steele_Nat_Nano_09}.
High spatial control and resolution of the electrostatic doping of semiconducting nanotubes will allow the realization of electronic and optoelectronic devices like diodes or phototransistors \cite{Sze} with tunable properties, which is not possible for devices based on chemically doped semiconductors. Moreover, a controlled confinement of single carriers in combination with a p-n junction \cite{Imamoglu_92} in a semiconducting nanotube could potentially enable future applications such as electrically driven single photon sources in the burgeoning field of carbon nanotube quantum optics \cite{Hogele_PRL_08}.

Here we report on a scanning photocurrent microscopy (SPCM) study of suspended semiconducting nanotube devices where the band profile is engineered by means of local gates in order to generate p-n junctions at controlled locations along the nanotube axis.

\begin{figure}[!b]
\includegraphics{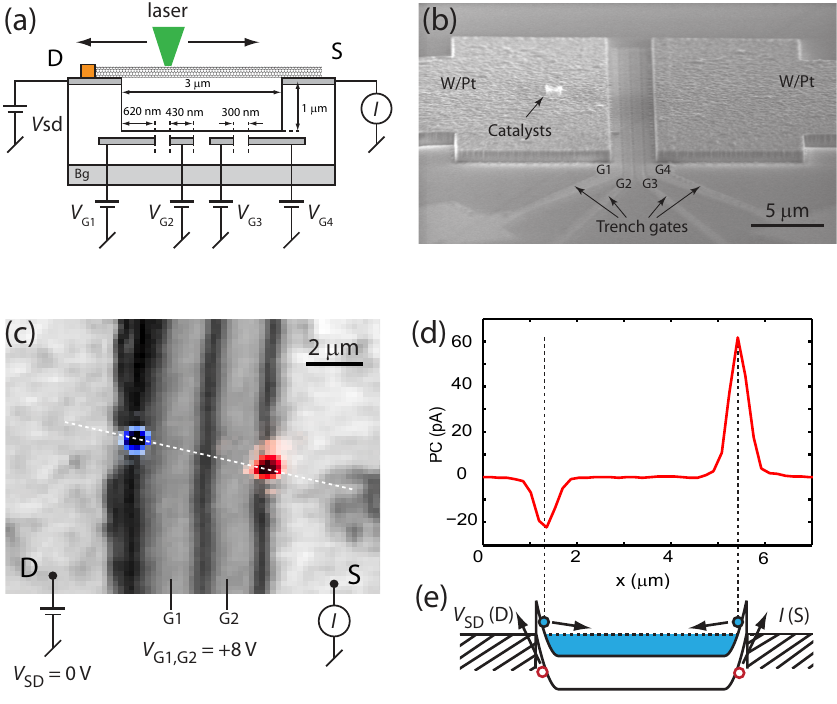}
\caption{\it{(a) Schematic of a device with four trench gates $G1-G4$. A diffraction-limited laser spot ($\lambda=532$nm) is scanned across the device and PC is recorded between source (S) and drain (D) contacts. (b) Scanning electron microscope image of a four trench gates device. (c) Superimposition of the PC image (red-blue scale) and the reflection image (grey scale) for a device with two trench gates separated by 250 nm, measured with $V_{G1}=V_{G2}=+8$ V and $V_{SD}=0$ V. A single semiconducting nanotube is highlighted with a dashed white line. (d) PC line profile recorded along the dashed line in panel (c) corresponding to the nanotube axis. (e) Corresponding band diagram with photogenerated carrier separation at the metal/nanotube interfaces.}}
\label{F1}
\end{figure}

\section{EXPERIMENT}

The devices consist of a nanotube grown between platinum electrodes over predefined trenches with a depth of 1 $\mu$m and widths of 3 or 4 $\mu$m. Up to four gates are defined at the bottom of the trenches. A schematic and a scanning electron microscopy image of a typical device with four gates and a 3 $\mu$m wide trench are shown in Figs.~\ref{F1} (a) and (b), respectively. The fabrication began with a p${++}$ silicon wafer used as a backgate covered by 285 nm of thermal silicon oxide. On top of this, gate electrodes made of 5/25 nm W/Pt were defined using electron-beam lithography, followed by the deposition of a 1100 nm thick SiO$_{2}$ layer. A 1000 nm deep trench was dry etched, leaving a thin oxide layer on top of the gates. A 5/25 nm W/Pt layer was then deposited to serve as source and drain contacts, and nanotubes were grown at the last fabrication step at a temperature of 900 $^\circ$C from patterned Mo/Fe catalysts \cite{Kong_Nat_98,Steele_Nat_Nano_09}. 

\begin{figure}[!b]
\includegraphics{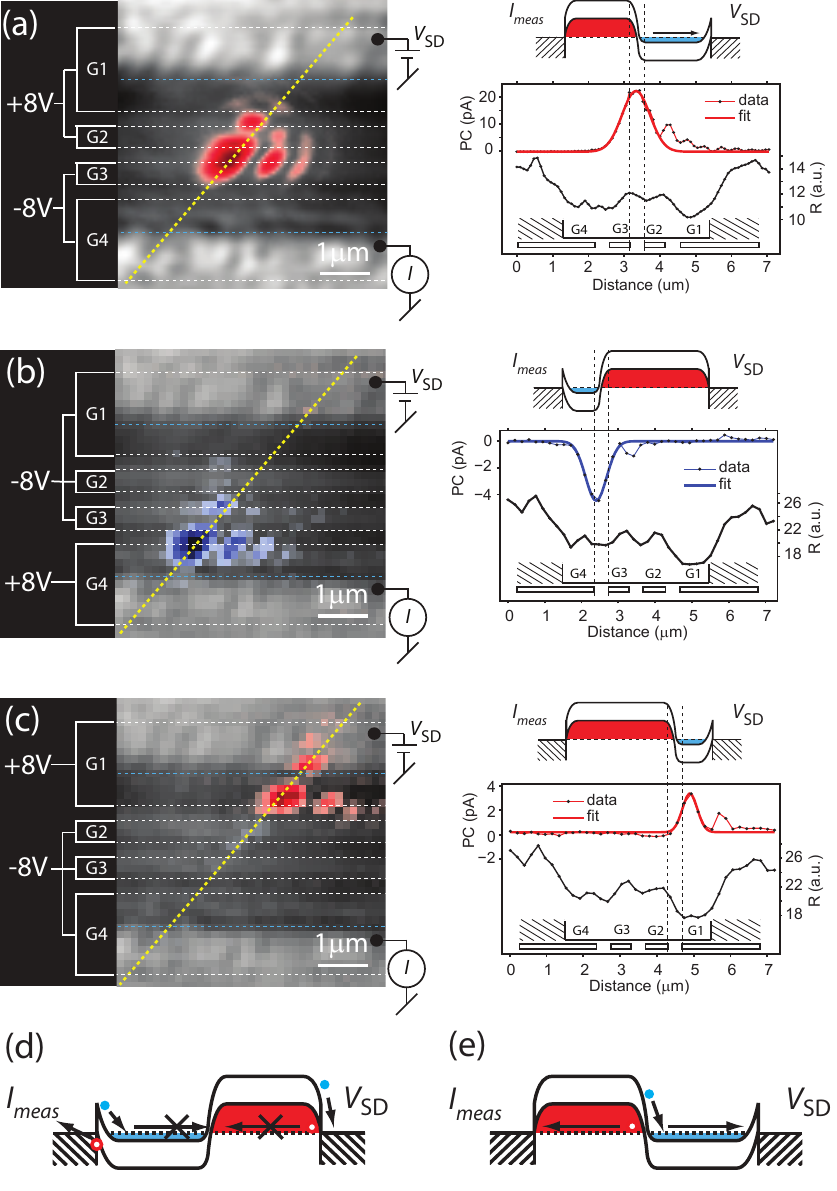}
\caption{\it{Tuning the position and polarity of a p-n junction using local gates. (a)-(c) Left column: superimposition of the PC and reflection images and the configurations of the potentials applied to the trench gates. The yellow dashed line corresponds to the nanotube axis. Right column: Corresponding band diagrams with PC and reflection intensities (R) recorded along the yellow dashed line as well as the position of the gates. Each PC line profile is fitted with a Gaussian discarding the diffraction-induced patterns. (d)-(e) Band diagrams illustrating the behavior of the photogenerated carriers at different positions: metal/nanotube interfaces and depletion region, respectively.}}
\label{F2}
\end{figure}

In SPCM, photocurrent (PC) is recorded as a laser spot is scanned across a sample. PC appears when photogenerated electrons and holes are separated by local electric fields in the device, such as those present at metal/nanotube interfaces \cite{Ahn_NL_07,Balasubramanian_NL05}, defect sites \cite{Balasubramanian_NL05} or p-n junctions \cite{Gabor_Science_09}. Our SPCM setup consists of a confocal microscope with a $NA=0.8$ objective illuminated by a $\lambda=532$ nm laser beam. The diffraction limited spot is scanned using a combination of two galvo-mirrors and a telecentric lens system while the dc PC signal and the reflected light intensity are recorded simultaneously in order to determine the absolute position of the detected PC features. Typical light intensities of 3 kW/cm$^{2}$ are used in this work.

\section{RESULTS AND DISCUSSION}

Fig.~\ref{F1} (c) shows the superimposition of the PC (blue-red scale) and reflection (gray scale) images of a device with two gates labeled G1 and G2 separated by 250 nm in a 4 $\mu$m wide trench, measured in vacuum ($\approx 10^{-4}$ mbar) at room temperature. The applied voltages are $V_{G1}=V_{G2}=+8$ V and $V_{SD}=0$ V. The PC image in combination with the measured transfer characteristics reveal the presence of a single p-type semiconducting nanotube crossing the trench, whose axis is indicated by a white dashed line. Fig.~\ref{F1} (d) shows the PC line profile recorded along the dashed line in panel (c) with two PC spots at the edge of the trench revealing the local electric field generated at the metal/nanotube interface. The corresponding band diagram is depicted in (e) with an illustration of the photogenerated carrier separation at the contacts. The asymmetry in the PC is due to different resistances for carriers at the source and drain contacts, for instance here a thinner Schottky barrier for electrons at the drain contact (D).\cite{Ahn_NL_07} 

\begin{figure*}[!t]
\includegraphics{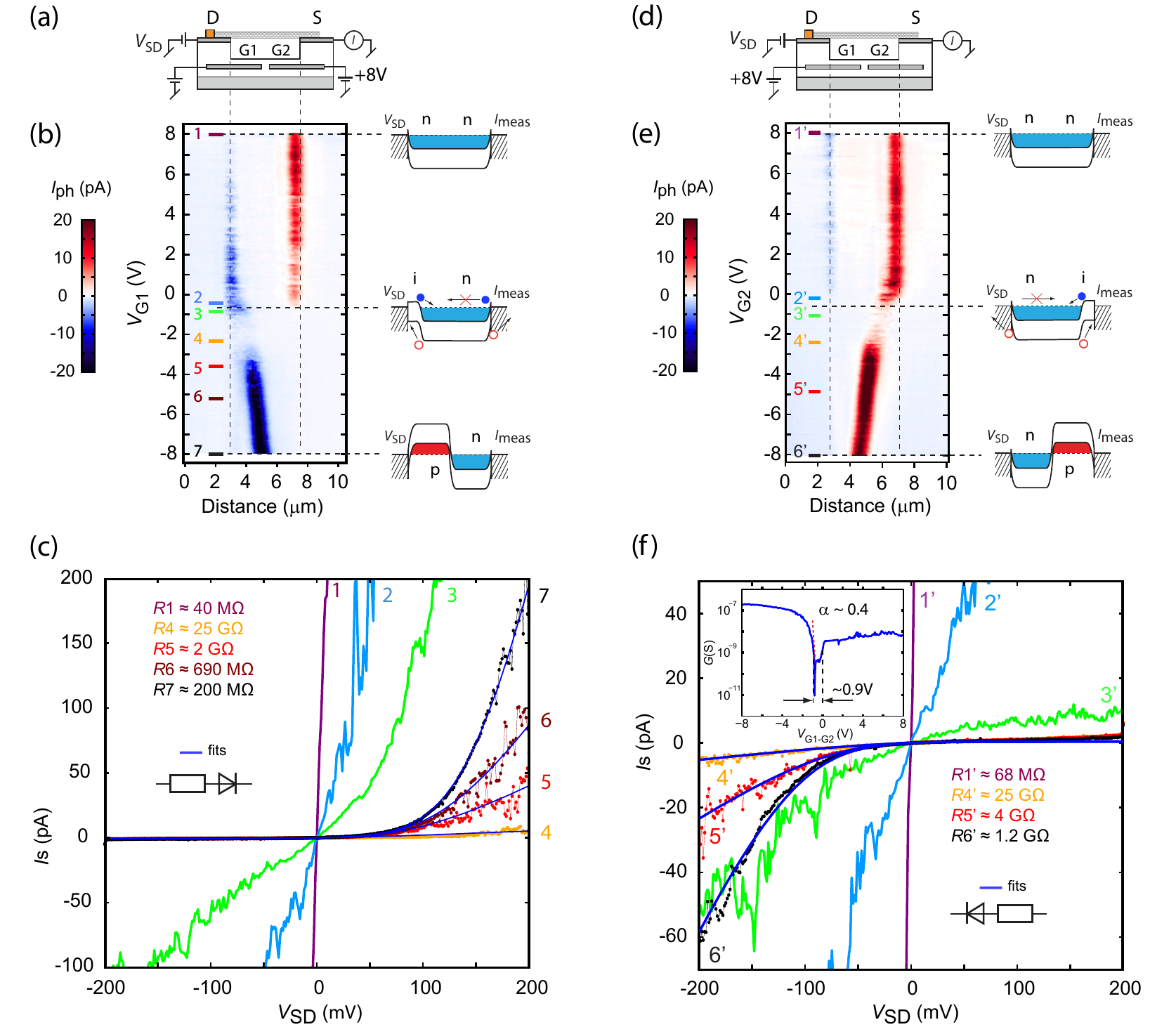}
\caption{\it{Imaging the emergence of a p-n junction. (a) Schematic of a device with two trench gates. G2 is set to $+8$ V and G1 is swept. (b) PC transition map recorded along the nanotube axis, with corresponding band diagrams for n-n, i-n and p-n regimes. (c) $I-V$ curves recorded at values of $V_{G1}$ labeled 1-7 in (b). Inset: values of the series resistances R1 for the linear n-n regime and R4-R7 for the rectification regime fitted with the Shockley diode model (blue curves). (d) G1 is set to $+8$ V and G2 is swept. (e) PC transition map recorded along the nanotube axis, with corresponding band diagrams for n-n, n-i and n-p regimes. (f) $I-V$ curves recorded at values of $V_{G2}$ labeled 1$'$-6$'$ in (e). Inset bottom right: values of the series resistances R1$'$ for the linear n-n regime and R4$'$-R6$'$ for the rectification regime fitted with the Shockley diode model (blue curves). Inset up left: Transfer characteristics of the device, where Vg corresponds to the voltage applied simultaneously to G1 and G2. $\alpha$ is the estimated gate efficiency.}}
\label{F3}
\end{figure*}

In Fig.~\ref{F2}, we demonstrate the imaging of a p-n junction, whose position and polarity can be tuned with the local gates. We use a four trench gate device similar to the one shown in Fig.~\ref{F1} (b). The measurements are performed at room temperature in air. With the same potential applied to all four gates, two PC spots appear at the trench edges with polarities depending on the gates potential, similar to those shown for the two gate device in Fig.~\ref{F1} (c). For opposite potentials (+8 V/-8 V) applied to groups of adjacent gates with configurations G1-G2,G3-G4 (panel (a)), G1-G2-G3,G4 (panel (b)) and G1,G2-G3-G4 (panel (c)), p-n junctions are created and clear PC spots appear at the electric field maxima corresponding to depletion regions. The images demonstrate that we can both move the position of the pn-junction and change its polarity using the local gates. Note that the metal/nanotube interface does not show PC signals due to the potential barrier formed at the depletion region that blocks one of the photogenerated carriers, as illustrated in panel (d). The patterns around the maximum intensity PC spots are due to diffraction effects from the structure of the gates. Gaussian fits to the PC signals along the nanotube axis (dashed yellow lines) show that the center of the depletion regions is positioned close to the center of the spacing between two gates with opposite potentials. 

In Fig.~\ref{F3} we study a two trench gates device illustrated in the schematic in Fig.~\ref{F3} (a) (4 $\mu$m wide trench and gate separation 250 nm). The measurements have been performed at room temperature in vacuum. A single semiconducting nanotube was found to cross the trench establishing an electrical contact between source and drain electrodes. Using the technique described in Refs. \cite{Rosenblatt_NL_02,Ahn_NL_07}, we estimate the bandgap to be $E_{g} \approx$ 400 meV, corresponding to a diameter of about 1.7 nm \cite{Odom98}, and find that the Fermi level at the contacts lies at about one third of the bandgap below the conduction band \cite{com1}. A transition from a fully n-type or n-n channel to p-n (n-p) configuration is studied in panels (a)-(c) ((d)-(f)) by applying a constant potential of $+8$ V to G2 (G1) and sweeping G1 (G2) from $+8$ V to $-8$ V. For each value of $V_{G1}$ ($V_{G2}$), the laser spot is scanned along the nanotube axis and the PC is recorded, Fig.~\ref{F3} (b) (Fig.~\ref{F3} (e)). For the range $ V_{G1} (V_{G2}) \geq 0$ V, the PC shows two contributions at the Schottky barriers. Below 0 V the negative (positive) PC signal starts to move its position towards the center of the trench and the positive (negative) PC signal vanishes.

Both effects are due to a transition of the band profile from pure n-n to the configuration depicted with the label i-n (n-i) where the drain (D) (source (S)) side of the n channel begins to pinch off and prevents electrons generated at the source (S) (drain (D)) Schottky barrier from reaching the drain (source) contact. The negative (positive) PC signal continues to move into the trench until it is suppressed below $V_{G1}=-1$ V ($V_{G2}=-0.8$ V) and then recovers around $V_{G1}=-3$ V ($V_{G2}=-2$ V). This low PC intensity likely indicates a shallow potential profile in which the electric field is not large enough to separate the photogenerated carriers. 
At $V_{G1}\approx -3$ V ($V_{G2}\approx -2.5$ V), the PC signal increases drastically up to about -22 pA (30 pA) and shifts slowly towards the center of the trench at $V_{G1} (V_{G2}) =-8$ V. This strong PC signal is the consequence of hole doping of the drain (source) side of the device, resulting in a p-n (n-p) junction with a large electric field in its depletion region \cite{Ahn_NL_07}, depicted in the band diagram corresponding to $V_{G1} (V_{G2}) =-8$ V in panel (b) (panel (e)).

In addition to PC imaging, we also perform $I-V$ measurements (dark current) for values of $V_{G1}$ ($V_{G2}$) indicated by labels 1-7 (1$'$-6$'$) in panel (b) (panel (e)). A progression from ohmic regime at $V_{G1} (V_{G2}) =+8$ V with a measured resistance of about 40 M$\Omega$ (68 M$\Omega$)\cite{note:one} to a clear rectification behavior starting below $V_{G1} (V_{G2})=-2$ V corresponding to $I-V$ curves 4-7 (4$'$-6$'$) with the forward current increasing with $\left|V_{G1}\right|$ ($\left|V_{G2}\right|$) is observed. The rectification curves 4-7 ((4$'$-6$'$)) can be fitted well to the Shockley diode model $I=I_{0} (e^{V_{SD}/(n \cdot V_{T})}-1)$ with a series resistor \cite{Lee_APL_05}, 
\begin{equation}
I=I_{0}\left[ \frac{n V_{T}}{I_{0} R} W \left( \frac{I_{0} R}{n V_{T}} e^{\frac{V_{SD}+I_{0} R}{n V_{T}}} \right)-1 \right] 
\end{equation}
where $I_{0}$ is the saturation current at reverse bias, $n$ is the ideality factor, $V_{T}$ is the room temperature thermal voltage of 26 mV, $R$ is the series resistance, $W$ is the Lambert $W$-function and $V_{SD}$ is the source-drain voltage. For a measured saturation current of about $I_{0}=4\cdot 10^{-13}$ A ($I_{0}=-4\cdot 10^{-13}$ A), we find the best fit with $n=1$ and $R4=25$ G$\Omega$, $R5=2$ G$\Omega$, $R6=690$ M$\Omega$ and $R7=200$ M$\Omega$ ($R4'=25$ G$\Omega$, $R5'=4$ G$\Omega$ and $R6'=1.2$ G$\Omega$). The decreasing value of $R$ with $\left|V_{G1}\right|$ ($\left|V_{G2}\right|$) is in good agreement with the band profiles model depicted on the right side of panel (b) (panel (e)), implying a decrease in width of the tunneling barrier for hole injection in the segment of the nanotube located above G1 (G2) when $\left|V_{G1}\right|$ ($\left|V_{G2}\right|$) increases. The higher current in forward bias for the p-n configuration compared to n-p is due to an asymmetry in the resistance at the source and drain contacts. We note that the estimated bandgap from the turn-on voltage ($V_{SD}\approx150$ mV) is not consistent with the value estimated from the transfer characteristic ($E_{g}\approx400$ mV). In addition to the systematic error from the estimation of the bandgap from the transfer characteristic using the method of Refs.~\cite{Rosenblatt_NL_02,Ahn_NL_07}, this difference can also potentially be due to diffusion effects for the carriers, recently observed by another group~\cite{Michigan_11}. Such discrepancies in our devices will be the subject of future investigations.

\section{CONCLUSIONS}

In summary, we have demonstrated the control of the position and polarity of a p-n junction in multigate suspended carbon nanotube devices. We created a p-n junction from a purely n-type channel and imaged its formation using SPCM. In the electrical characteristics, a corresponding transition is observed from the linear resistance of a transistor channel to the non-linear rectification of an ideal diode. The high degree of control of ideal p-n junctions using the local gates, combined with the precise photocurrent imaging of the p-n junction position, demonstrate the potential of carbon nanotubes and the SPCM technique in optoelectronics applications.

\begin{acknowledgments}

This research was supported by a Marie Curie Intra European Fellowship within the 7th European Community Framework Programme, a FOM projectruimte, and NWO Veni and Vidi programs.

\end{acknowledgments}


\begin{thebibliography}{19}
\bibitem [{com()}]{com1}%
  \BibitemOpen
  \href@noop {} {}\bibinfo {note} {For this device, a quantitative estimation
  of the bandgap out of the transfer characteristic is only possible in vacuum.
  Indeed, under atmospheric conditions, a water-oxygen layer present on the
  SiO$_{2}$ will screen the gates at positive voltage and prevent an electron
  current to flow in the nanotube (see Ref.\cite{Aguirre_AM_09}).}\BibitemShut
  {Stop}%
\bibitem [{not()}]{note:one}%
  \BibitemOpen
  \href@noop {} {}\bibinfo {note} {The difference in the resistance here is due
  to a combination of hysteresis and inaccuracy in the linear fit due to too
  few measurement points in this narrow $V_{SD}$ range.}\BibitemShut {Stop}%
\end{thebibliography}

%

\end{document}